\documentclass[apj]{emulateapj}
\usepackage{graphicx}
\usepackage{amssymb}
\usepackage{amsmath}
\usepackage{lineno}
\usepackage{listings_test}
\usepackage{color}
\usepackage{natbib}
\usepackage[labelsep=none,font=small]{caption_test}
\definecolor{deepblue}{rgb}{0,0,0.5}
\definecolor{deepred}{rgb}{0.6,0,0}
\definecolor{deepgreen}{rgb}{0,0.5,0}
\definecolor{lightgray}{rgb}{0.9,0.9,0.9}

\DeclareFixedFont{\ttb}{T1}{txtt}{bx}{n}{9} 
\DeclareFixedFont{\ttm}{T1}{txtt}{m}{n}{9}  

\newcommand\pythonstyle{\lstset{
language=Python,
basicstyle=\ttm,
otherkeywords={self, XDGMM,fit,bic_test,ShuffleSplit,validation_curve,array,condition,nan,sample,read_model,component_test,fit_model,get_local_SB,get_logR,append,get_SN},             
keywordstyle=\ttb\color{deepblue},
emph={MyClass,__init__,some_data,labels},          
emphstyle=\ttb\color{deepred},    
stringstyle=\color{deepgreen},
frame=tb,                         
showstringspaces=false            %
backgroundcolor=\color{lightgray},%
}}

\lstnewenvironment{python}[1][]
{
\pythonstyle
\lstset{#1}
}
{}

\lstdefinestyle{pythonstyle}{
language=Python,
basicstyle=\ttm,
otherkeywords={self, XDGMM,fit,bic_test,ShuffleSplit,validation_curve,array,condition,nan,sample,read_model,component_test,fit_model,get_local_SB,get_logR},             
keywordstyle=\ttb\color{deepblue},
emph={MyClass,__init__,some_data,labels},          
emphstyle=\ttb\color{deepred},    
stringstyle=\color{deepgreen},
frame=tb,                         
showstringspaces=false            %
backgroundcolor=\color{lightgray},
}

\makeatletter
\newcommand{\Spvek}[2][r]{%
  \gdef\@VORNE{1}
  \left(\hskip-\arraycolsep%
    \begin{array}{#1}\vekSp@lten{#2}\end{array}%
  \hskip-\arraycolsep\right)}

\def\vekSp@lten#1{\xvekSp@lten#1;vekL@stLine;}
\def\vekL@stLine{vekL@stLine}
\def\xvekSp@lten#1;{\def\temp{#1}%
  \ifx\temp\vekL@stLine
  \else
    \ifnum\@VORNE=1\gdef\@VORNE{0}
    \else\@arraycr\fi%
    #1%
    \expandafter\xvekSp@lten
  \fi}
\makeatother

\newcommand{\emp}{{\sc empiriciSN}}
\newcommand{\xd}{{\sc XDGMM}}

\def\myx{\mathbf{x}}
\def\mymu{\mathbf{\mu}}
\def\mysigma{\mathbf{\Sigma}}
\def\mylambda{\mathbf{\Lambda}}
\newcommand{\seq}{\,\!{=}\,\!} 

\begin{document}



\title{EmpiriciSN: Re-sampling Observed Supernova/Host Galaxy Populations using an XD Gaussian Mixture Model}

\author{{Thomas W.-S. Holoien}\altaffilmark{1,2,3,4}, {Philip~J.~Marshall}\altaffilmark{1,2}, {Risa~H.~Wechsler}\altaffilmark{1,2}}

\altaffiltext{1}{Kavli Institute for Particle Astrophysics and Cosmology, Department of Physics, Stanford University, Stanford, CA, 94305}
\altaffiltext{2}{SLAC National Accelerator Laboratory, Menlo Park, CA, 94025}
\altaffiltext{3}{Department of Astronomy, The Ohio State University, 140 West 18th Avenue, Columbus, OH 43210, USA}
\altaffiltext{4}{Center for Cosmology and AstroParticle Physics (CCAPP), The Ohio State University, 191 W. Woodruff Ave., Columbus, OH 43210, USA}



\begin{abstract}

We describe two new open source tools written in Python for performing extreme deconvolution Gaussian mixture modeling (XDGMM) and using a conditioned model to re-sample observed supernova and host galaxy populations. {\xd} is new program for using Gaussian mixtures to do density estimation of noisy data using extreme deconvolution (XD) algorithms that has functionality not available in other XD tools. It allows the user to select between the {\sc AstroML} \citep{vanderplas12,ivezic15} and \citet{bovy11} fitting methods and is compatible with {\sc scikit-learn} machine learning algorithms \citep{pedregosa11}. Most crucially, it allows the user to condition a model based on the known values of a subset of parameters. This gives the user the ability to produce a tool that can predict unknown parameters based on a model conditioned on known values of other parameters. {\emp} is an example application of this functionality that can be used for fitting an XDGMM model to observed supernova/host datasets and predicting likely supernova parameters using on a model conditioned on observed host properties. It is primarily intended for simulating realistic supernovae for LSST data simulations based on empirical galaxy properties.

\end{abstract}






\section{Introduction}
\label{sec:intro}

The problem of inferring a distribution function given a set of samples from that distribution function and the problem of finding overdensities in this distribution function are common issues in many areas of science, particularly astronomy \citep[e.g.,][]{skuljan99,hogg05,bovy12}. Gaussian mixture models (GMMs), which model an underlying density probability distribution function (pdf) using a sum of Gaussian components, are a commonly used tool for solving density estimation problems such as these \citep{ivezic14}. However, traditional GMMs do not have the ability to incorporate measurement noise into the density calculation, and often in astronomy we must deal with observations that have multiple sources of noise with very different properties. For problems such as this, the ``extreme deconvolution'' GMM (XDGMM) technique must be used.

XD was originally outlined by \citet{bovy11}, and provides a way to perform Bayesian estimation of multivariate densities modeled as Gaussian mixtures \citep{ivezic14}. XDGMMs have already proven useful for modeling underlying distributions using noisy observations for multiple areas of astronomy, from velocity distributions of nearby stars \citep{bovy09,bovy10} to photometric redshifts and quasar probabilities of SDSS sources \citep{bovy12} to 3-D kinematics of stars in the Sagittarius stream \citep{koposov13}. However, the potential of XD models to be used as predictive tools has yet to be explored. An XDGMM is able to model the complicated correlations between various parameters in a many-dimensional dataset. If this model could be conditioned on the known values of some of these parameters, it could be used to predict likely values for the remaining parameters, allowing the sampling of realistic properties without knowledge of how the various parameters are correlated. While there are multiple existing implementations of the XDGMM algorithm, no existing tool has this functionality.

One potential use of an XDGMM prediction tool is to sample likely parameters for a supernova (SN) given the parameters of a host galaxy. The problem of supernova simulation is a common one for large-scale sky surveys, as it is useful both for planting fake supernovae (SNe) in existing data to test detection efficiency for calculating SN rate \citep{melinder08,graur14} and for creating realistic simulated data to test data processing pipelines of upcoming surveys, such as the Large Synoptic Survey Telescope \citep[LSST;][]{ivezic08}.  For various applications it can also be useful to place realistic simulated SN within a realistic galaxy distribution in a cosmological context, for example to understand the connection between observational biases in host detection and cosmological observables.
In each of these cases, having realistic SN properties is essential for avoiding the introduction of further uncertainty or biasing detection of new sources. Further, many known correlations between host and SN properties are based on physical quantities, such as host mass, metallicity, and star formation rate, that must be inferred from observations using theoretical models, introducing further uncertainty \citep[e.g.,][]{sullivan10,childress13,graur16a,graur16b}. An XDGMM model trained only on a wide range of empirical, observed host properties could be used to sample realistic supernova properties without the need for theoretical models, removing this as a source of uncertainty. However, in order to build such a tool, a new implementation of XDGMM that allows for the conditioning of the model is needed.

The rest of this paper is laid out as follows. In \S\ref{sec:XDGMM} we describe the {\xd} class\footnote{https://github.com/tholoien/XDGMM} \citep{holoien16a} and the new functionality that differentiates it from existing XDGMM fitting tools. In \S\ref{sec:empiricisn} we describe the {\emp} supernova prediction tool\footnote{https://github.com/tholoien/empiriciSN} \citep{holoien16b} and demonstrate its functionality. Finally in \S\ref{sec:disc} we summarize the capabilities of our software and describe some of the preliminary results obtained using \emp.


\section{XDGMM}
\label{sec:XDGMM}

As described in \S\ref{sec:intro}, XDGMM fitting methods are useful tools for performing density estimation of noisy data, a situation that occurs often in astronomy. When we began our research into building a tool to predict the properties of supernovae based on observed host galaxy properties, XDGMM modeling seemed to be a natural way to use machine learning to fit the underlying distributions of the numerous host and supernova properties in our dataset. Furthermore, by conditioning such a model on known host properties, we can create a model based solely on the supernova properties of interest, and sample from this conditioned model to predict supernova properties for a given host. However, existing tools, such as {\sc AstroML}\footnote{http://www.astroml.org/index.html} \citep{vanderplas12,ivezic15} and the {\sc Extreme-Deconvolution} tool\footnote{https://github.com/jobovy/extreme-deconvolution} from \citet{bovy11}, provided XD fitting methods but did not have the ability to condition the model that we required. In addition, though the {\sc AstroML} implementation of XDGMM utilizes some of the functionality of the {\sc scikit-learn} GMM class \citep{pedregosa11}, neither tool implements the {\sc scikit-learn} algorithms that could be used to perform cross-validation (CV) tests to optimize model parameters. Because the ability to condition and perform cross-validation on XDGMM models could be useful for other astronomical studies, we decided to first build our own implementation of XDGMM that provides access to both the {\sc AstroML} and {\sc Extreme-Deconvolution} fitting methods and implements the functionality we needed to build a supernova fitting tool and make it available to the public.


\subsection{Fitting Methods}
\label{sec:fit_methods}

Both the {\sc AstroML} and \citet{bovy11} fitting algorithms are able to successfully fit a Gaussian Mixture Model using extreme deconvolution, and improving or editing their methods was not one of our goals with this project. However, as the two tools use slightly different algorithms to perform fits, we provided the ability for the user to select between the two when using \xd. Brief descriptions of each method are provided here.

{\sc AstroML} is an open source Python module for machine learning and data mining and provides a wide range of statistical and machine learning tools for analyzing astronomical datasets \citep{vanderplas12,ivezic15}. One of the provided tools is an implementation of XD Gaussian Mixture Modeling, and is based on the algorithms described in \citet{bovy11}. Though slower than the \citet{bovy11} {\sc Extreme-deconvolution} tool, which makes use of {\sc OpenMP} for parallelizing the model fitting process, the {\sc AstroML} implementation of XDGMM provides a clean user interface similar to that of the GMM implementation of {\sc scikit-learn} \citep{pedregosa11}, and we based our interface for {\xd} on that of the {\sc AstroML} tool. We also use the {\sc AstroML} implementation of several of the methods used to score datasets under an existing model.

The utility of extreme deconvolution for density estimation of astronomical datasets was first described in \citet{bovy11}, and the {\sc Extreme-Deconvolution} tool provided by the authors of that manuscript was one of the first tools to implement XDGMM fitting methods. Though Python, R, and IDL wrappers are available, {\sc Extreme-Deconvolution} is built in C and uses {\sc OpenMP} to parallelize the fitting method. As such, it provides a significantly faster fit than the {\sc AstroML} XDGMM tool.

Though the {\sc Extreme-Deconvolution} provides faster performance, we elected to make the {\sc AstroML} fitting method the default fitting method of {\xd} for two reasons. First, the {\sc AstroML} implementation provides more stable fit results, and is less prone to issues resulting from outlying data than the \citet{bovy11} tool. Second, as a Python module, {\sc AstroML} is easily installable on most systems, while {\sc Extreme-Deconvolution} requires a more detailed installation, as C compilers and the availability of {\sc OpenMP} vary from system to system. Because of this, while {\sc AstroML} is required for installing {\xd}, {\sc Extreme-Deconvolution} is not, and is only imported if the user attempts to perform a fit using the \citet{bovy11} method.

Listing~\ref{lst:fit_sample} shows an example of how to create a new \xd{} object, fit a model to a dataset, and sample data from the model. The fitting and the sampling interface was purposefully built to mimic the {\sc AstroML} XDGMM interface so that our {\xd} class could be substituted for theirs in existing code.

\begin{python}[
caption={---An example of the fitting interface for \xd. We purposefully built this to use the same interface as the {\sc AstroML} XDGMM tool.},
label={lst:fit_sample},
frame=single,
backgroundcolor=\color{lightgray}]
from xdgmm import XDGMM
xd = XDGMM()
X, Xerr = (data, errors)
xd.fit(X, Xerr)
xd.sample(2000)
\end{python}


\subsection{Component Selection}
\label{sec:component_selection}

When fitting a Gaussian Mixture Model to a dataset, it is necessary to choose the number of Gaussian components to use in the model. If the number of components in the model is too small, the model will be too simplistic, and will underfit the data, but if the number of components is too large, the model will be too flexible and will overfit the data. In either case, a subsequent sample drawn from the model will not accurately represent the dataset used to train the model. In addition, with large datasets and large numbers of parameters being fit, the computation costs will rapidly become expensive, resulting in very long fit times. Thus, it is important to choose a number of components that can fit the data well without overfitting and which doesn't place unnecessary stress on computational resources.

\subsubsection{Bayesian Information Criterion}
\label{sec:bic}

One method for choosing the correct number of components for the model is to use the Bayesian Information Criterion \citep[BIC;][]{schwarz78}. The formula for the BIC is given in Equation~\ref{eq:bic} below.

\begin{equation}
\label{eq:bic}
\textrm{BIC} = -2\log{\hat{L}} + k\log{n}
\end{equation}

Here, $\hat{L}$ is equal to the maximized likelihood function of the model being scored, $k$ is the number of free parameters being estimated (e.g., the number of components in a GMM), and $n$ is the number of observations used to fit the model. If the BIC is calculated for a number of different models that each use a different number of free parameters, the one with the lowest BIC is the preferred model. Since it incorporates both a likelihood score and a component that incorporates the number of free parameters, it penalizes models with too many degrees of freedom, which can help avoid overfitting.

Our {\xd} class computes the BIC score for the current model using a specific dataset in the same way that the {\sc scikit-learn} GMM class computes the BIC, except that our class can also account for uncertainty on the input data when computing the BIC. (If the user inputs a covariance matrix with a dataset, the {\xd} class will incorporate these uncertainties into the model covariance matrix before calculating the BIC.) We have also provided a function that can compute the BIC for a given dataset for a range of numbers of components, allowing the user to compare different models and select the ideal one. This functionality is demonstrated in Listing~\ref{lst:bic}.

\begin{python}[
caption={---A demonstration of the BIC test function, which allows the user to compare different {\xd} models with different numbers of components.},
label={lst:bic},
frame=single,
backgroundcolor=\color{lightgray}]
param_range = np.array([1,2,3,4,5,6,7,8,9,10])
bic, optimal_n_comp, lowest_bic =
    xd.bic_test(X, Xerr, param_range)
\end{python}

\begin{figure}
\centering\includegraphics[width=0.95\linewidth]{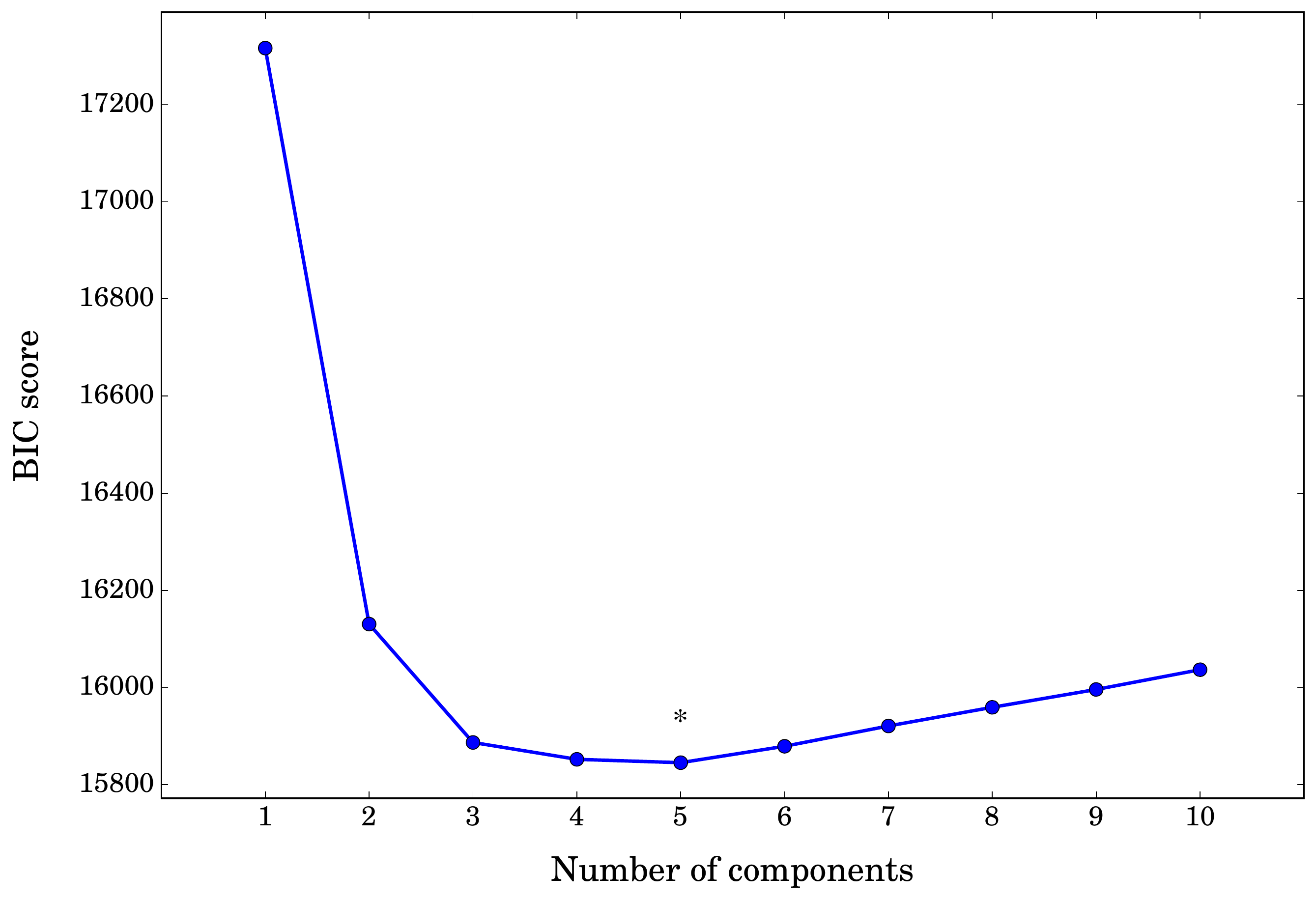}
\caption{BIC scores computed by {\xd} using the {\sc AstroML} XDGMM demo dataset \citep{vanderplas12} for different models with the number of components ranging from 1 to 10. The optimal number of model components based on the BIC is 5.}
\label{fig:bic}
\end{figure}

In the example code of Listing~\ref{lst:bic}, the {\xd} object computes the BIC score for the data and uncertainties contained in the {\sc X} and {\sc Xerr} arrays for a number of components ranging from 1 to 10. It then returns the {\sc bic} array, which contains the score for each model, the optimal number of components (defined as the number of components in the model with the lowest BIC score), and the lowest BIC score. These results can be used for direct comparison, or can be plotted to see the results visually. In Figure~\ref{fig:bic}, we show the BIC results for the {\sc AstroML} XDGMM demo dataset \citep{vanderplas12}. The minimum BIC value occurs with 5 Gaussian components in the model, indicating that models with higher numbers of components overfit the data.

\subsubsection{Machine Learning}
\label{sec:machine_learning}

An alternative way to determine the number of components is to perform a cross-validation test with a range of numbers of components. In order to allow the user to perform such a test, we have made the {\xd} class a subclass of the {\sc scikit-learn} {\sc BaseEstimator} class and implemented all the {\sc scikit-learn} functions necessary for the standard {\sc scikit-learn} cross-validation tools. We use the mean log-likelihood of a dataset under the given model as the score for cross-validation. Because {\xd} extends {\sc BaseEstimator}, a cross-validation test can be performed by simply passing a {\xd} object and a dataset into the {\sc scikit-learn} cross-validation functions. A demonstration of computing a validation curve using the same {\sc AstroML} demo dataset for 1 to 10 components is given in Listing~\ref{lst:cv}.

\begin{python}[
caption={---A demonstration of a cross-validation test performed using the {\sc scikit-learn} {\sc validation\_curve} function.},
label={lst:cv},
frame=single,
backgroundcolor=\color{lightgray}]
param_range = np.array([1,2,3,4,5,6,7,8,9,10])
shuffle_split = ShuffleSplit(len(X), 3,
                 test_size=0.3)
train_scores,test_scores =
    validation_curve(xd, X=X, y=Xerr,
                     param_name=
                         "n_components",
                     param_range=param_range,
                     n_jobs=3, 
                     cv=shuffle_split)
\end{python}

It is important to note is the trick used to pass errors to the {\sc scikit-learn} methods. Normally for unsupervised learning you would only pass an X array, and not a ``target'' y array, to the {\sc validation\_curve} method. However, an error array must be passed to the {\xd} {\sc fit} method, and {\sc validation\_curve} simply uses the y parameter as the second argument for the {\sc fit} function. Thus, by treating the error array as the ``target'' array, we can pass it to our {\sc fit} function so that it can be used in fitting the data.

\begin{figure}
\centering\includegraphics[width=0.95\linewidth]{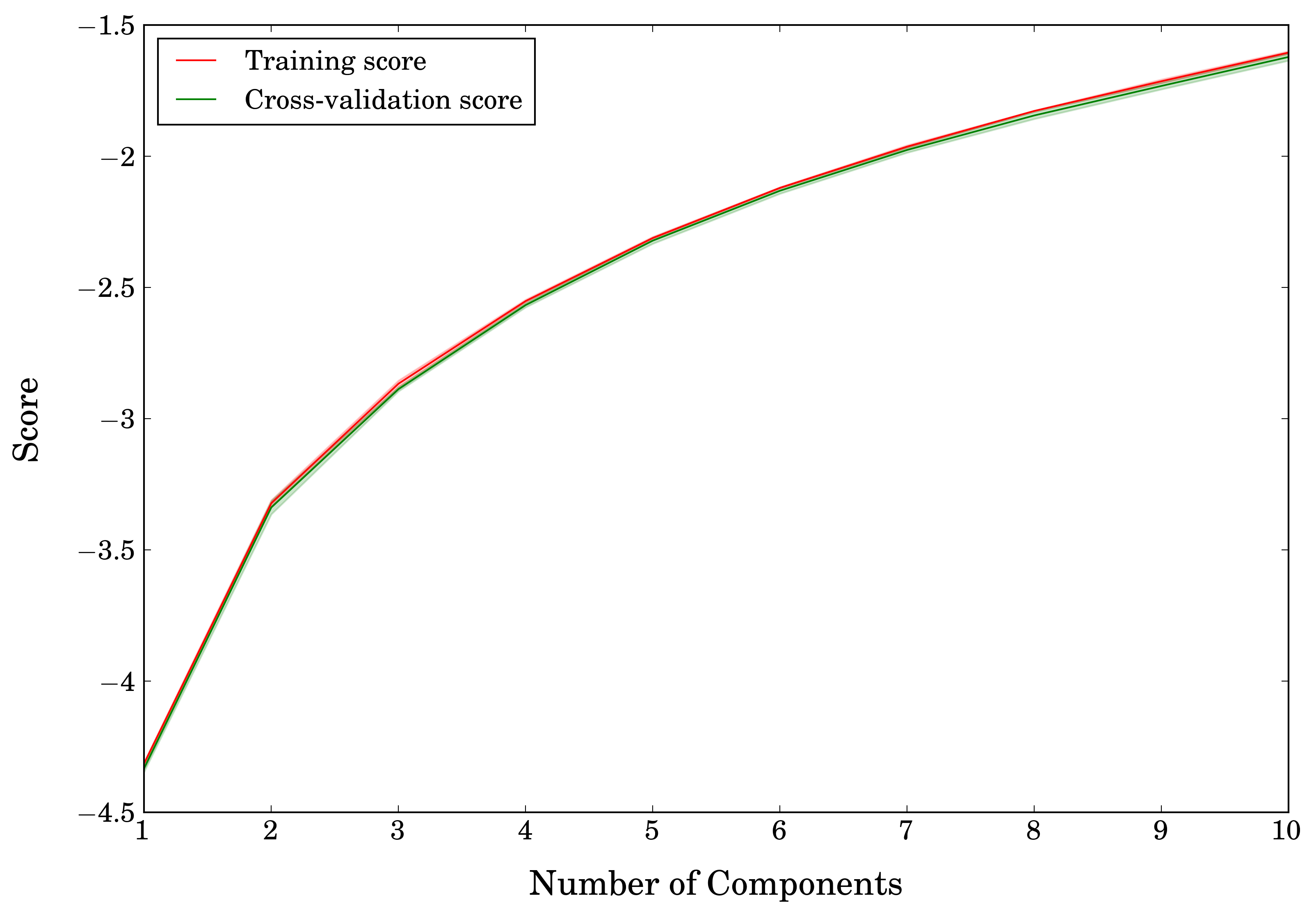}
\caption{Results of the cross-validation test performed in Listing~\ref{lst:cv} on the {\sc AstroML} demo dataset \citep{vanderplas12}. The red line shows the mean training scores for each number of components in the test, and the green line shows the mean scores for the test sample. The shaded regions indicate the standard deviation for each. The cross-validation test prefers a maximum number of components for this dataset.}
\label{fig:val_curve}
\end{figure}

Figure~\ref{fig:val_curve} shows the results of the cross-validation test above. The cross-validation test prefers the maximum number of components (10) that we allowed for the model (and in fact, the score continues to rise as more components are added beyond 10). This is a result of the particular dataset being fit: the likelihood of the data being fit increases with more components, and there is enough structure to the data that even with a large number of Gaussians, the trained model continues to be a good predictor of new data. However, increasing the number of components in the model rapidly causes the fit algorithm to become computationally expensive, especially for the astroML algorithm. While a model with a large number of components may be mathematically superior, in most cases the BIC seems to provide a way to find a model that is ``good enough'' to fit the data well, while also keeping the number of components at a value that keeps the computation of new fits reasonable. We recommend trying both tests with a given dataset to see if the results differ substantially before settling on a choice for the number of components to use when fitting a model.

Once we know the optimal number of components to use, it is straightforward to fit a model to the data (see Listing~\ref{lst:fit_sample}). Figure~\ref{fig:val_curve} replicates the results of the {\sc AstroML} Extreme Deconvolution example \citep{vanderplas12} using our {\xd} code. We first create a ``true'' distribution and a ``noisy'' distribution using the \citet{vanderplas12} demo code, then we fit a model to the noisy distribution using 5 Gaussian components and sample 2000 data points from the model. We can see that even with only 5 components, the distribution sampled from the XDGMM model is able to replicate the true data sample despite being fit using the ``measured'' noisy distribution. This demonstrates why extreme deconvolution is such a powerful tool for modeling datasets such as this.

\begin{figure}
\centering\includegraphics[width=0.98\linewidth]{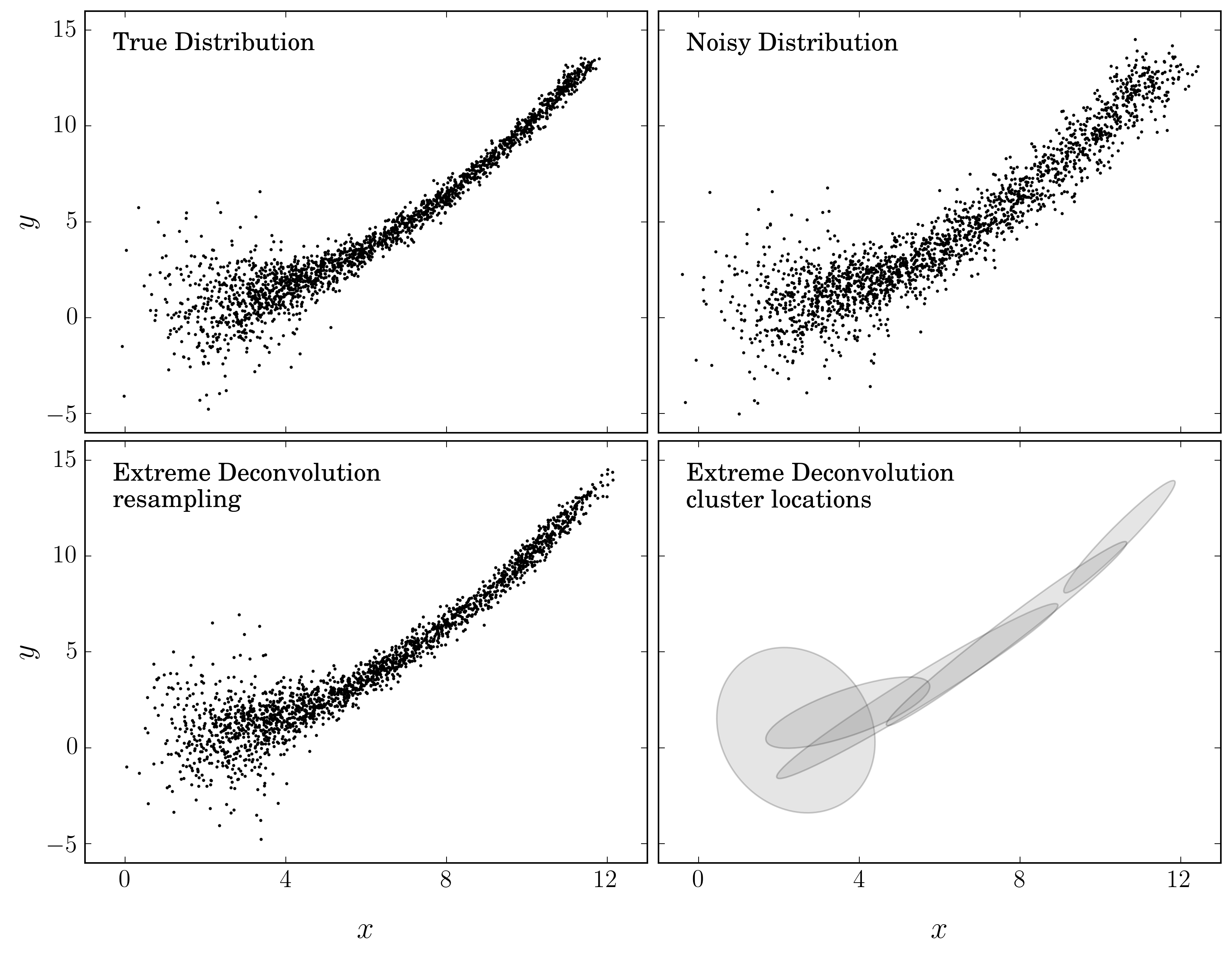}
\caption{Replication of the results of the {\sc AstroML} Extreme Deconvolution example \citep{vanderplas12}. \emph{Top Left:} the ``true'' distribution of the demo dataset. \emph{Top right:} the ``noisy'' distribution of the demo dataset, meant to simulate observed data points. \emph{Bottom right:} The 5 Gaussian components from the {\xd} model after fitting the model to the data. \emph{Bottom left:} 2000 data points sampled from the {\xd} model. The resampled dataset closely matches the true distribution, despite being fit using the noisy distribution.}
\label{fig:sample_dist}
\end{figure}


\subsection{Conditioning the Model}
\label{sec:conditioning}

A primary motivation behind this implementation of extreme deconvolution was to develop a tool that can be used to predict model parameters based on known values of other parameters. To do this, the model must be conditioned on the known parameter values, after which we can then sample values of the non-conditioned parameters from the conditioned model. Neither the {\sc AstroML} nor the \citet{bovy11} implementations of XDGMM contain this functionality, and we have implemented it in our software.

First, we briefly discuss the mathematics of a conditional Gaussian mixture model. The probability distribution for a Gaussian mixture with $K$ components is given by \citep{bishop06,rasmussen06}:

\begin{equation}
\label{eq:pdf}
p\left(\myx\right) = \sum_{k=1}^{K} \pi_k \mathcal{N}\left( \myx \mid \mymu_k,\mysigma_k\right)
\end{equation}

Here, $\pi_k$, $\mymu_k$, and $\mysigma_k$ are the mixing coefficient (weight), means, and covariances of the $k$-th Gaussian component. Given the jointly Gaussian vectors $\myx_A$ and $\myx_B$ and the above Gaussian mixture, we have:

\begin{equation*}
\myx = \Spvek{\myx_A;\myx_B},\quad
\mymu_k = \Spvek{\mymu_{kA};\mymu_{kB}},
\end{equation*}
\begin{equation*}
\mysigma_k = \Spvek{\mysigma_{kAA}\; \mysigma_{kAB};\mysigma_{kBA}\; \mysigma_{kBB}},\quad
\mylambda_k = \mysigma_k^{-1}
\end{equation*}

The conditional distribution of $\myx_A$ given $\myx_B$ for the $k$-th Gaussian component is given by \citep{bishop06}:

\begin{equation}
\label{eq:cond_pdf}
\begin{split}
p_k\left(\myx_A \mid \myx_B\right) &=
\frac{p_k\left(\myx_A, \myx_B\right)}{p_k\left(\myx_B\right)}\\
&=\mathcal{N}\left( \myx_A \mid \mymu_{kA\mid B},\mylambda_{kAA}^{-1}\right)
\end{split}
\end{equation}

The conditional mean vector of the $k$-th Gaussian component is given by:

\begin{equation}
\label{eq:cond_means}
\mymu_{kA \mid B} = \mymu_{kA} - \mylambda_{kAA}^{-1}\mylambda_{kAB} \left(\myx_B-\mymu_{kB}\right)
\end{equation}

Finally, the $k$-th conditional mixing coefficient is given by:

\begin{equation}
\label{eq:cond_weight}
\pi_k'=\frac{\pi_k \mathcal{N}\left( \myx_B \mid \mymu_{kB},\mysigma_{kBB}\right)}{\sum_{k}\mathcal{N}\left( \myx_B \mid \mymu_{kB},\mysigma_{kBB}\right)}
\end{equation}

Thus, the conditional probability distribution for the whole GMM is given by:

\begin{equation}
\label{eq:cond_pdf_full}
p\left(\myx_A \mid \myx_B\right) = \sum_{k=1}^K \pi_k'p_k\left(\myx_A \mid \myx_B\right)
\end{equation}

The resulting conditioned GMM has the same number of Gaussian components as the original GMM, but has fewer dimensions, given by the number of dimensions in $\myx_A$. We can then use this conditioned model to sample values for the parameters in $\myx_A$ given the known quantities in $\myx_B$.

When using XDGMM as a prediction tool for astronomical quantities, it may often be the case that the user wants to condition the model on parameter measurements that have measurement uncertainties. Conditioning a model on a particular value of $\myx_B\seq\myx_{B,0}$ is equivalent to marginalizing out $\myx_B$ assuming a delta function PDF for it. If we include uncertainties on the measurement of $\myx_B$ in the form of a covariance matrix $C_B$, we can incorporate these uncertainties into the conditioning by adding $C_B$ to the covariance array of the unconditioned GMM prior to conditioning the model. The result will still be a GMM, but its components will be i) broader, since the extra covariance $C_B$ will end up being added to the component covariance matrix $\mylambda_{kAA}^{-1}$, and ii) weighted differently, since the weights in Equation~\ref{eq:cond_weight} are themselves functions of $\myx_B$.

We have built two different but equally straightforward interfaces for conditioning the model. In order to condition the model, the {\xd} object needs to be informed which of the parameters of the model (e.g., $x$ and $y$ in the sample dataset) have values on which to condition the model. The {\xd} object stores the parameters in a specific order based on their order in the dataset that was used to fit the model---for example, $x$ is stored first and $y$ second for the demo dataset used here, since the data was passed to the model as $(x,y)$ pairs. In some cases, such as this simple demo, it may be easy for the user to remember the order of the parameters, and we have built one conditioning interface to take advantage of such cases. Listing~\ref{lst:cond1} demonstrates this first interface. In this conditioning method, the user passes one or two arrays to the {\sc condition} function: one array containing values for each parameter (either a value for conditioning or {\sc NaN} if the parameter is not being used for conditioning), and an optional second array containing uncertainties on the parameter values.

\begin{python}[
caption={---A demonstration of the first interface for model conditioning the model using the demo dataset and the known value $y=1.5\pm0.05$. In this method, the user knows the indices of the parameters to use for conditioning and passes arrays for the parameter values and uncertainties to the {\sc condition} function.},
label={lst:cond1},
frame=single,
backgroundcolor=\color{lightgray}]
fixed_X = np.array([np.nan, 1.5])
unc = np.array([0.0,0.05])
new_xd = xd.condition(X_input=fixed_X,
                      Xerr_input=unc)
\end{python}

However, we recognize that in many cases, the user may be fitting large datasets with many parameters, and maintaining the proper order for conditioning may be difficult. For this reason, the {\xd} object also allows the user to label the parameters in the model. If the labels have been set, the user can then use a dictionary object which links the labels for parameters to condition the model on with (value, uncertainty) pairs to condition the model. In Listing~\ref{lst:cond2} we demonstrate this functionality. Here we label the parameters `x' and `y' and then pass a dictionary containing only a value for $y$ to the {\sc condition} function. The conditioned model that results will be the same regardless of the method used for conditioning; the different interfaces are simply supplied so that the user can choose whichever method they prefer.

\begin{python}[
caption={---A demonstration of the second interface for model conditioning. Labels for the different parameters in the model are first set by the user, and then a dictionary object is used for conditioning.},
label={lst:cond2},
frame=single,
backgroundcolor=\color{lightgray}]
xd.labels = np.array(['x','y'])
fixed = {'y':(1.5,0.05)}
new_xd2 = xd.condition(X_dict=fixed)
\end{python}

\begin{figure}
\centering\includegraphics[width=0.98\linewidth]{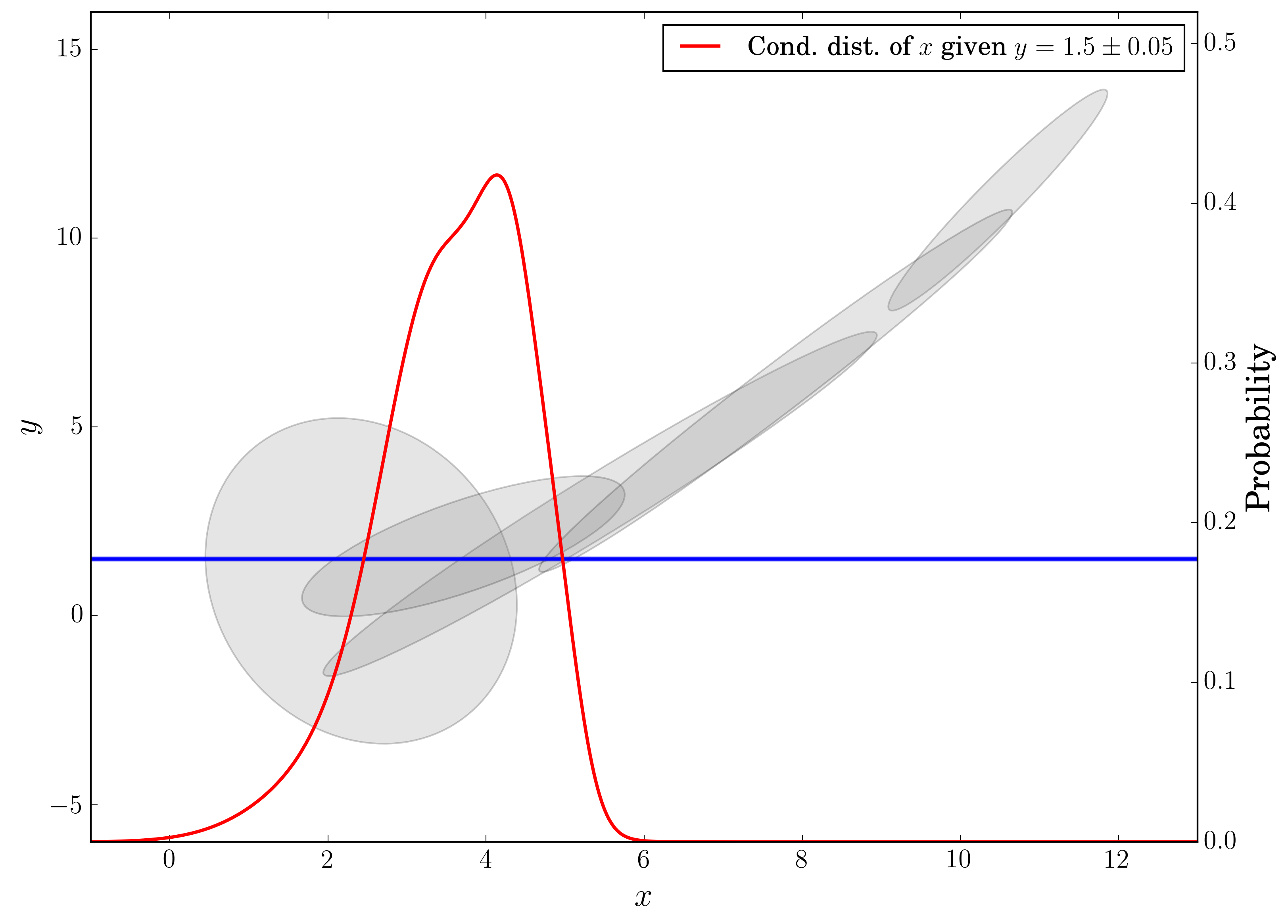}
\caption{Probability distribution of $x$ from the demo model conditioned on $y=1.5\pm0.05$. The gray ellipses show the 5 components of the original {\xd} model (the bottom-right panel of Figure~\ref{fig:sample_dist}), the blue line shows the value of $y$ used to condition the model, and the red line shows the resulting distribution of $x$ (right scale). Note how conditioning the model changes the weights and means of the components of the model, and how the measured value of $y$ essentially rules out several model components.}
\label{fig:cond_model}
\end{figure}

Once the model has been conditioned, the resulting model will have the same number of components as the original model but will have fewer dimensions, as it is now a model only for the parameters that were not conditioned out. In the code above, we have conditioned the demo model based using $y=1.5\pm0.05$, and the resulting model is a 5 component GMM for the $x$ parameter. Figure~\ref{fig:cond_model} shows the resulting probability distribution of $x$. Conditioning the model changes the weights, means, and covariances of the components of the model. The constraint on $y$ essentially rules out several components of the original model, significantly reducing their weight in the conditioned model.

\begin{figure}
\centering\includegraphics[width=0.98\linewidth]{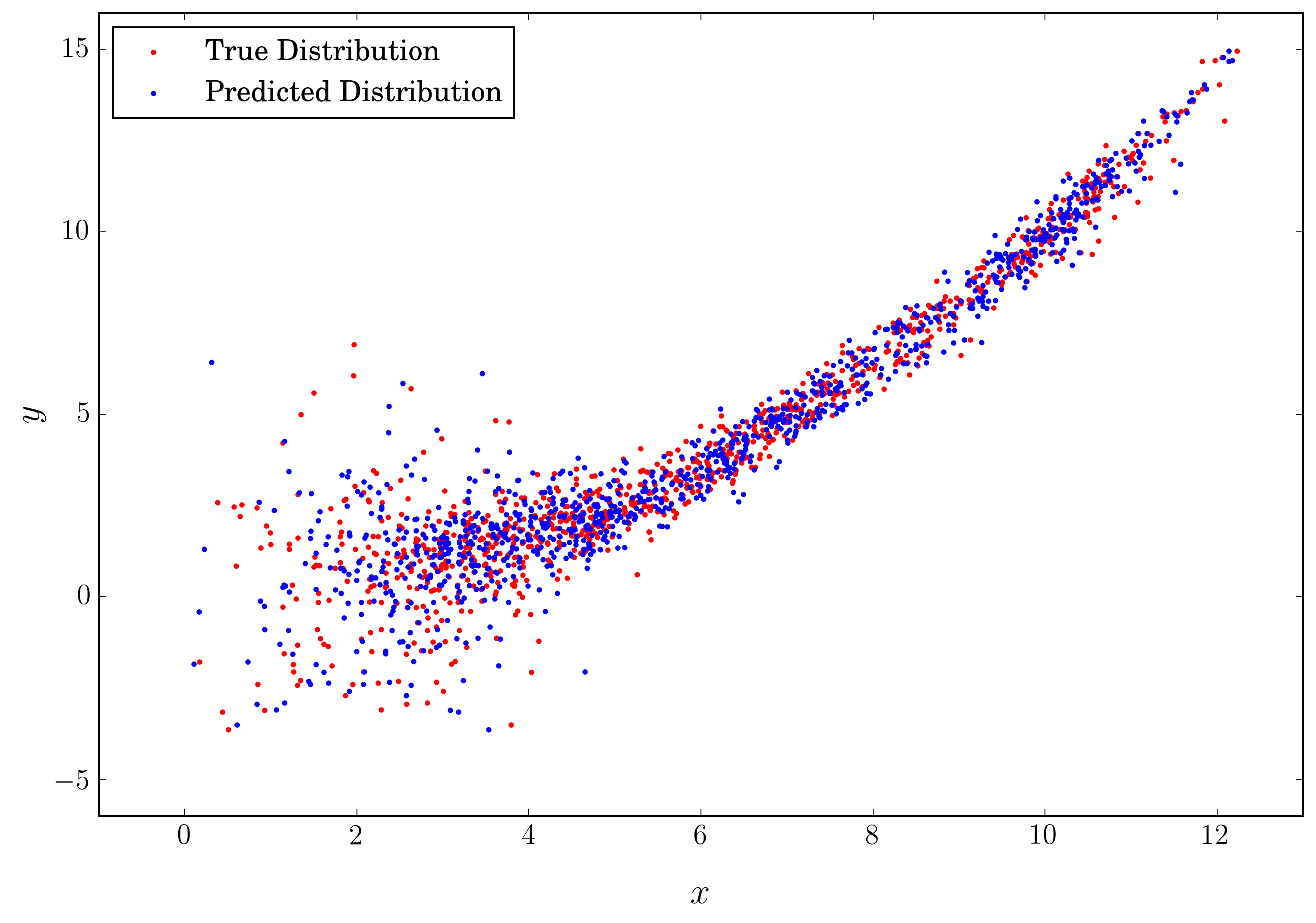}
\caption{Distribution of ``observed'' $(x,y)$ pairs sampled from the demo {\xd} model (red) compared with $(x,y)$ pairs that predict an $x$ value for each observed $y$ value (blue). Though a predicted $x$ value for a given $y$ value may not exactly match the true $x$ value, the overall distribution is the same, which indicates that the {\xd} model functions properly as a prediction tool.}
\label{fig:cond_pred}
\end{figure}

As stated before, one potential use of a conditioned model is to create a ``prediction engine'' that can predict some parameters using an XDGMM model conditioned on known values of other parameters. To demonstrate this, we samplde 1000 data points from our original, unconditioned model to create a dataset to be compared with our predictions. This represents a new ``observed'' dataset for the $x$ and $y$ parameters. Now if we had only observed the $y$ values from this dataset and wanted to predict a likely $x$ value for each $y$ value, we can condition the model on each of these $y$ values in turn and draw a predicted $x$ value from the conditioned model. In reality, {\xd} would likely be used to predict values for parameters that have not been measured, so this provides a good way to test whether the tool is functioning properly---these predicted $x$ values should follow the same distribution of the observed $x$ values---and Figure~\ref{fig:cond_pred} shows that this is the case. Though the predicted $x$ for a single given $y$ value may not match the observed $x$ value, the fact that the overall distributions match indicates that {\xd} provides an accurate prediction tool.

\section{{\sc EmpiriciSN}}
\label{sec:empiricisn}

It is well established that supernova properties, such as light curve color and shape, have a dependence on their host environments \citep[e.g.,][]{modjaz08,sullivan10,childress13,galbany14,graur16a,graur16b}. Planting simulated supernovae in real or simulated imaging data has a variety of uses---for instance, fake supernovae can be used to test the detection efficiency of supernova searches, which is a necessary quantity to know for calculating supernova rates, and simulated data including supernovae can be used to train data processing and automated source detection pipelines for future surveys like LSST. Modeling the distribution of supernovae within a cosmologically motivated galaxy distribution can also be used to test how the connection between their properties and the underlying structure can impact various observables.

It is important to ensure that supernovae simulated for these purposes have properties that accurately match their environments and are consistent with each other, otherwise the results of these studies wouldn't be applicable to real-world conditions. However, many of the known supernova-host correlations are based on physical parameters of the host---e.g., star formation rate, mass, metallicity---which must be inferred using theoretical models from observations \citep[e.g.,][]{sullivan10,childress13,graur16a,graur16b}. Though many of these theories are fairly well-established, this introduces additional uncertainty into the selection of supernova properties.

In this section we outline {\emp}, a tool for predicting realistic Type Ia supernova (SN Ia) properties based only on observed host galaxy properties. Our goal in creating {\emp} was to provide a model trained on observed empirical host and supernova properties, thus eliminating the need for using theoretical models to infer the host galaxy's physical properties. As this requires calculating correlations between many supernova and host properties and the subsequent conditioning of the model, it provides a real-world use for our {\xd} class. Our default model is trained using supernova and galaxy properties that can be generated by {\sc SNCosmo} \citep{barbary14} and {\sc CatSim} \citep{connolly14} so that it can be used for generating realistic supernovae for LSST Twinkles\footnote{https://github.com/DarkEnergyScienceCollaboration/Twinkles} simulations (LSST DESC, in prep.). These include the {\sc SALT2} Type Ia light curve parameters ({\sc x0}, {\sc x1}, and c; \citealt{guy07}), the host redshift, the 10 rest-frame host colors obtainable with $ugriz$ magnitudes, the separation of the supernova from the host nucleus in units of the host effective radius, and the local surface brightness at the location of the supernova in all 5 $ugriz$ filters.


\subsection{Input Catalogs}
\label{sec:input_catalogs}

In order to model the 20 host and supernova properties listed above, we require a large dataset with consistent measurements of the {\sc SALT2} parameters and consistent host photometry. We decided to use a sample of supernovae taken from the Supernova Legacy Survey \citep[SNLS;][]{astier06,sullivan11} and the Sloan Digital Sky Survey \citep[SDSS;][]{york00} Supernova Survey to build a model for our data.

All SALT2 light curve parameters for the SNLS supernovae and a portion of the parameters for our SDSS supernovae are taken from the Joint Light-curve Analysis \citep[JLA;][]{betoule14}, a project to analyze light curves of supernovae discovered by SNLS, SDSS, and other sources. This includes 242 spectroscopically confirmed Type Ia supernovae from SNLS and 374 spectroscopically confirmed Type Ia supernovae from the SDSS SN survey. Because the JLA catalog provides a peak magnitude rather than the {\sc x0} SALT2 parameter, we used {\sc SNCosmo} to fit the JLA light curves ourselves assuming the redshifts, {\sc x1}, and c parameters provided, and calculated {\sc x0} in this way. In order to increase the size of our sample to be large enough to model, we also include the remaining spectroscopically confirmed SNe Ia from the SDSS SN search as well as the photometric SNe Ia with a spectroscopic host redshift from \citep{sako14}. We include the photometric SNe Ia because doing so provides us with an additional 906 SNe, though we acknowledge that some of these may not actually be Type Ia. \citet{sako14} provide the {\sc x0} SALT2 parameter, but since they use a slightly different model than the JLA model, we correct the \citet{sako14} {\sc X0} values by a factor of $10^{(0.108)}$ to make them consistent with the values measured from the JLA light curves.

We then searched the coordinates of all host galaxies from our SN samples in the SDSS Data Release 12 \citep[DR12;][]{alam15} and obtained model magnitudes, model fluxes, effective radii, and morphology for all galaxies that were detected in DR12 data. This reduced our sample to 159 supernovae from SNLS and 1273 supernovae from SDSS. For the purposes of calculating local surface brightness, all galaxies in the sample are considered to have either an exponential or a de Vaucouleurs surface brightness profile, and we use whichever model is considered more likely by the SDSS pipeline in the $r$-band as the model for each galaxy. The $r$-band effective radius of the best-fit surface brightness profile was used to convert the separation from the host nucleus measured in \citet{betoule14} and \citet{sako14} from arcseconds to units of $R/R_e$.

The host galaxy model magnitudes are magnitudes generated for each filter assuming the same best-fit surface brightness profile and incorporate a convolution with the image PSF to account for PSF effects. The magnitudes have been corrected for Galactic extinction and corrected to rest-frame using {\sc kcorrect} \citep{blanton07} before being used to calculate host colors. The model magnitudes are ideal for unbiased galaxy color measurements, and while other datasets may not measure galaxy magnitudes and colors in the same way, we believe they should still be able to obtain reasonable results from our default model. 

Though we have taken steps to make our SN and host galaxy sample as consistent as possible, we acknowledge that both SN datasets were obtained after applying selection criteria, particularly the JLA sample, and we have not incorporated any corrections into our model to correct for these selection effects. Thus the distribution probability of SN parameters obtained by {\emp} combines both the true underlying distribution in nature and the detection efficiencies of each survey, and it may be necessary for the user to correct for the detection efficiency when using our default model to simulate SNe. We also caution that, though the dataset spans a wide range of SN and host galaxy properties, it is possible that certain galaxy types (e.g., low-mass galaxies) are underrepresented and therefore not well-modeled by the default model. We will be taking steps to quantify potential shortcomings of the default catalog in a future release of the software.

However, despite the survey effects present in our default datasets, when combined they still provide us with a fairly large dataset with a wide range of SN and host galaxy properties. Furthermore, we have built {\emp} to be easily updated with additional or different catalogs by the user; as long as the data files are formatted in a manner similar to our default dataset, an {\sc Empiricist} object can fit a model to any number of supernova and host properties. This allows the user to make adjustments to the provided SDSS/SNLS dataset, extend the model to include additional catalogs, or even to model different types of supernovae, in the future.

Testing using the BIC test described in \S\ref{sec:bic} indicated that the preferred number of Gaussian components for our dataset was 6, and we have provided a default 6-component model that is available with the {\emp} software. The default model also has built-in labels for each of the SN and host parameters so that these can be used for conditioning, as described in \S\ref{sec:conditioning}.


\subsection{Demo/Results}
\label{empiricisn_demo}

The {\emp} class object is called {\sc Empiricist}, and here we demonstrate some of its basic functionality. If using the default model fit to our SDSS and SNLS data sample, or another model that has already been fit to a dataset, it is straightforward to declare a new Empiricist object and read in the existing model, as shown in Listing~\ref{lst:emp1}. A model can be loaded upon creation of an {\sc Empiricist} object by passing the model file name to the {\sc model\_file} argument of the constructor. Alternatively, Listing~\ref{lst:emp2} demonstrates how to run a test for the optimal number of components and fit a new model to a dataset. The {\sc fit\_model} function always saves the newly fit model to a file, either with a the default name `empiriciSN\_model.fit' or with a file name passed into the function's {\sc filename} argument. After fitting a model, that model is stored in the {\sc Empiricist} object and can be used for SN prediction without the need to load the model again.

If using the default model, the local surface brightness at the location of the supernova in each filter are 5 of the host properties necessary for sampling a realistic SN for a host. However, the separation of the SN from the host nucleus is one of the properties being fit, and the local surface brightness cannot be calculated until a location has been chosen. Because of this, the prediction of a realistic SN proceeds in two steps: first, a position is selected based on a subset of host parameters, and then the local surface brightness is calculated and the SALT2 parameters of the SN are sampled from a model conditioned on the full range of host parameters.

\begin{python}[
caption={---An example of creating a new {\sc Empiricist} object and reading in our pre-computed default model. The {\sc Empiricist} object can also be created with the model already loaded by passing the file name to the {\sc model\_file} argument of the constructor.},
label={lst:emp1},
frame=single,
backgroundcolor=\color{lightgray}]
from empiriciSN import Empiricist
emp = Empiricist()
emp.read_model('default_model.fit')
\end{python}

In order to select a location for a SN, the user can condition on any subset of the parameters used to fit the model. The indices of the parameters to condition on must be passed as an argument to the {\sc get\_logR} function, as must the index of the $\log{R/R_e}$ parameter. Any indices may be used except the first three, which are assumed to be the three SALT2 parameters for the SN. The {\sc Empiricist} object will condition the XDGMM model using the data passed into the function and then return a value of $\log{R/R_e}$ sampled from the conditioned model. Uncertainties on the quantities being used for conditioning the model may be used, but are not required. Listing~\ref{lst:rad_select} demonstrates this functionality, using the host galaxy redshift and 10 host colors to condition the model and select a location for the SN.

\begin{python}[
caption={---A demonstration of the interface for testing for the optimal number of components to use for a dataset and fitting a model to that dataset. The {\sc component\_test} function's third argument is the range of values for the number of Gaussian components to test using the BIC test built into the {\xd} model; in this case we use the values $1-8$. The newly fit model is automatically saved to a file with the default name `empiriciSN\_model.fit', or with a name passed into the {\sc fit\_model} function's {\sc filename} argument.},
label={lst:emp2},
frame=single,
backgroundcolor=\color{lightgray}]
X, Xerr = (data, errors)
comp_range = np.array([1,2,3,4,5,6,7,8])
bics, optimal_n_comp, lowest_bic =
    emp.component_test(X=X, Xerr=Xerr,
                       component_range=
                           comp_range)
emp.fit_model(X=X, Xerr=Xerr,
              n_components=optimal_n_comp)
\end{python}

\begin{python}[
caption={---A demonstration of selecting a location for a SN given a subset of host parameters using our default model. In this case, the indices used for conditioning represent the redshift and the 10 $ugriz$ colors of the host. The X and Xerr arrays represent arrays of the host redshift and colors and uncertainties on these quantities, respectively.},
label={lst:rad_select},
frame=single,
backgroundcolor=\color{lightgray}]
X, Xerr = (data, errors)
cond_ind =
    np.array([3,5,6,7,8,9,10,11,12,13,14])
logR = emp.get_logR(cond_indices=cond_ind,
                    R_index=4, X=X, Xerr=Xerr)
\end{python}

When using our default model, the {\sc get\_local\_SB} function can be used to select a local surface brightness at the location of the SN once a location has been sampled for a given host. The host data passed into this function must be an array of 21 surface brightness parameters. The first index should be the host Sersic index (1 for an exponential profile, 4 for a de Vaucouleurs profile), and this should be followed by sets of magnitude, magnitude uncertainty, effective (half light) radius in arcseconds, and radius uncertainty for each of the 5 $ugriz$ filters. The function returns two arrays, one with the local surface brightness in units of magnitudes per square arcsecond for each filter and the other containing the uncertainties on the local surface brightnesses. The magnitudes are assumed to be K-corrected and corrected for Galactic extinction, as this is what the default model was trained on. Listing~\ref{lst:local_SB} demonstrates this.

\begin{python}[
caption={---A demonstration of calculating the local surface brightness of the host, if using our default model. The {\sc SB\_params} array is assumed to be of the form described in the text. The logR value is assumed to have been fit using {\sc get\_logR} as demonstrated in Listing~\ref{lst:rad_select}.},
label={lst:local_SB},
frame=single,
backgroundcolor=\color{lightgray}]
SB_params = [params_for_host]
SB, SB_err =
    emp.get_local_SB(SB_params=SB_params,
                     R=logR)
\end{python}

Finally, once we have a location and local surface brightnesses, it is straightforward to sample a SN for the host galaxy using the {\sc get\_SN} function. In Listing~\ref{lst:get_sn} we demonstrate how to select a single SN for a given host. The resulting array contains the SALT2 {\sc x0}, {\sc x1}, and c parameters for the sampled SN.

\begin{figure}
\centering\includegraphics[width=0.98\linewidth]{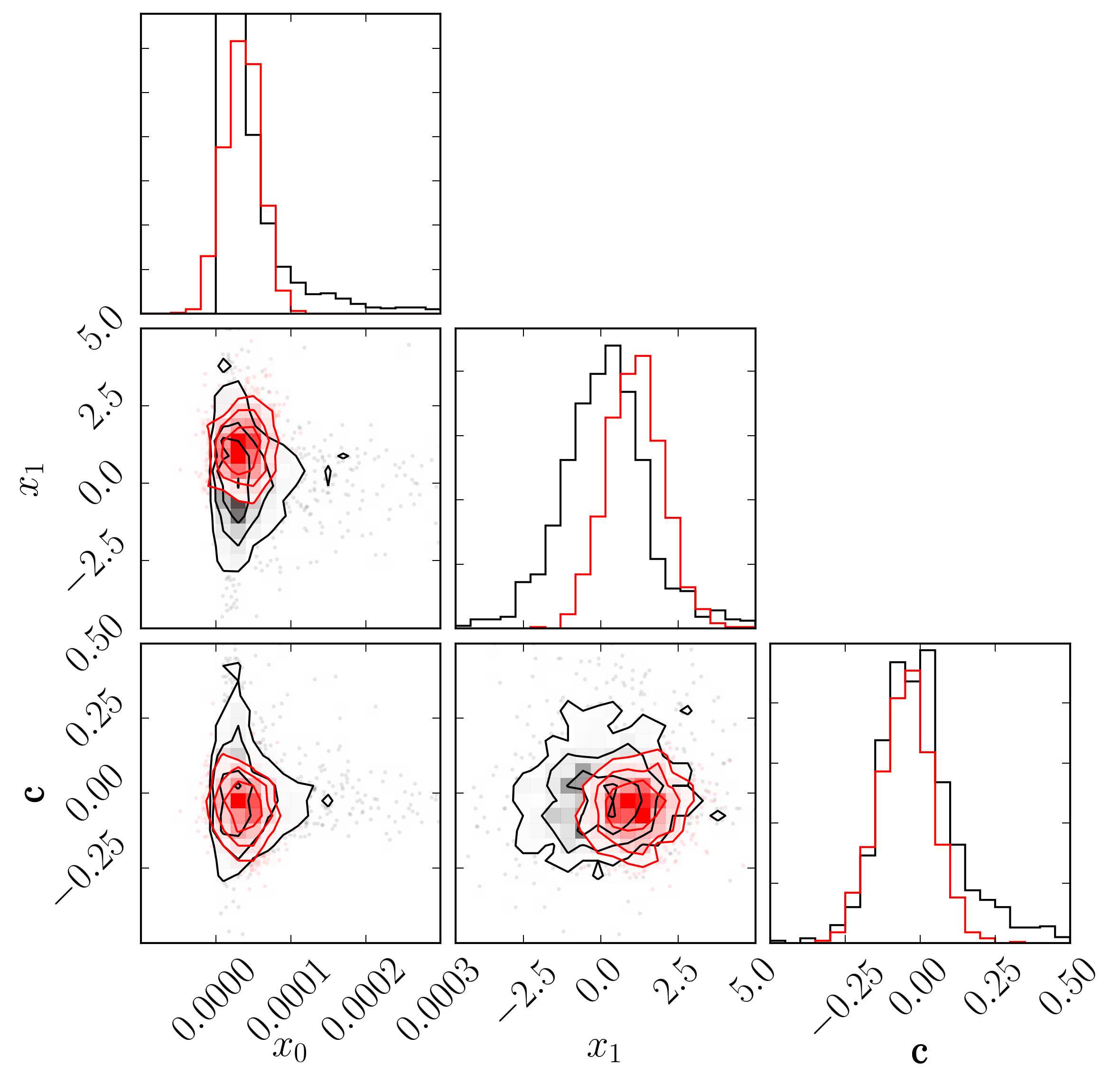}
\caption{Distributions of the three SALT2 SN parameters measured in our full dataset (black) compared to the parameters for 1000 SNe drawn using {\emp} for a single set of host galaxy parameters. The SN location and local surface brightness were also sampled using {\emp}, as described in Listings~\ref{lst:rad_select} and \ref{lst:local_SB}. The distributions of the various properties can be very different for a single position in a single host than what they are for the full sample.}
\label{fig:corner_plot}
\end{figure}

\begin{python}[
caption={---The supernova selection interface for the {\sc Empiricist} object. Here we combine the radius and the host properties fit in the previous two Listings with our initial magnitude and redshift arrays to create an array with all the host properties needed to condition the model and sample supernova properties. Here we are using the array index version of {\xd} conditioning as opposed to the dictionary version.},
label={lst:get_sn},
frame=single,
backgroundcolor=\color{lightgray}]
host_params =
    np.array([np.nan,np.nan,np.nan])
host_params = np.append(X[0],logR)
host_params = np.append(host_params,X[1:])
host_params = np.append(host_params,SB)
host_err =
    np.append(np.array([0.0,0.0,0.0]),
              Xerr[0])
host_err = np.append(host_err, 0.0)
host_err = np.append(host_err, Xerr[1:])
host_err = np.append(host_err, SB_err)

predicted_SN = emp.get_SN(X=host_params,
                       Xerr=host_err,
                       n_SN=1)
\end{python}

Figure~\ref{fig:corner_plot} shows a plot of the measured SALT2 parameter distributions from our full SN sample in black and the distribution of parameters for 1000 SNe that were sampled for a single host galaxy in our sample, with location and local surface brightness calculated as described above. As can be seen in the Figure, the distributions for the various properties can be quite different for a single position in a single host than what they are for the full sample, which is the behavior we expect.

For a more detailed walkthrough of the {\emp} software, please see the demo notebook on the {\emp} github page.


\section{Discussion}
\label{sec:disc}

In this paper we have summarized the capabilities of two new pieces of Python software that implement Extreme Deconvolution Gaussian Mixture Modeling for density estimation in astrophysical contexts.

{\xd} is a new implementation of XDGMM that extends existing XD algorithms and has several new capabilities that are not present in any previously existing XDGMM implementation. It allows the user to choose between either the {\sc AstroML} \citep{vanderplas12,ivezic15} or the {\sc Extreme-Deconvolution} code of \citet{bovy11} for performing XDGMM fits to data, and uses the {\sc AstroML} interface for fitting and sampling so that existing code that uses {\sc AstroML} does not need to be modified. It extends the {\sc scikit-learn} {\sc BaseEstimator} class so that cross-validation methods will work, and also has a function for computing the Bayesian Information Criterion of a model given a certain dataset, in order to allow the user to test different model parameters. Finally, and most crucially, {\xd} models can be conditioned on a given subset of data, and the conditioned model can then be used to sample the remaining parameters. This allows our {\xd} class to function as a prediction tool, which will have a wide variety of astronomical applications. {\xd} is easily extendible to include additional implementations of XDGMM methods or even different types of fitting solutions altogether, such as a Dirichlet process GMM. As long as the basic distribution remains a multivariate Gaussian, our machine learning and conditioning algorithms should still function as intended.

{\emp} is an example use of {\xd} as a prediction tool and is designed to predict realistic supernova parameters for a given set of host galaxy parameters. We have built an XDGMM model based on 4 Type Ia supernova parameters (The SALT2 {\sc x0}, {\sc x1}, and color parameters and the location of the supernova in the host galaxy) and 16 host galaxy parameters (redshift, $ugriz$ colors, sersic index, and $ugriz$ local surface brightness at the location of the SN) that is trained on a dataset comprising 159 supernovae from SNLS and 1273 supernovae from SDSS. Given the redshift, $ugriz$ magnitudes, and surface brightness profiles for a galaxy, {\emp} can select a location of a supernova relative to the effective radius of the galaxy, compute the local surface brightness in all 5 $ugriz$ filters, and sample realistic SALT2 parameters for the supernova given the host properties using this default model. The user can also train a new model based on a new dataset, if the default dataset is not desired, or to change the parameters used in the model. This makes it capable of performing similar predictions for other types of supernovae, or performing simpler, catalog level simulations, leaving out the SN position and local surface brightness information. {\emp} was primarily built to be used for planting realistic supernovae in simulated survey data, and is already being implemented for this purpose to model LSST supernovae.  In the near future we also intend to combine this tool with realistic large area galaxy mock catalogs, with potential applications for various current and future surveys including DES, DESI, and LSST. 

Both {\xd} and {\emp} are open source projects. We welcome further contributions and collaboration from the community.

\section*{Acknowledgments}

We thank Rahul Biswas and Bob Nichol for useful discussion and feedback.
TW-SH is supported by the DOE Computational Science Graduate Fellowship, grant number DE-FG02-97ER25308. PJM and RHW acknowledge support from the U.S.\ Department of Energy under contract number DE-AC02-76SF00515.

Funding for SDSS-III has been provided by the Alfred P. Sloan Foundation, the Participating Institutions, the National Science Foundation, and the U.S. Department of Energy Office of Science. The SDSS-III web site is http://www.sdss3.org/.


\bibliographystyle{aasjournal}
\bibliography{bibliography.bib}

\end{document}